\def\selectedoptions{}
\ifx\empty\selectedoptions
  \def\selectedoptions{final}
\fi

\documentclass[
   \selectedoptions
  ]
  {aipproc}

\def\selectedlayoutstyle{6x9} 
\layoutstyle\selectedlayoutstyle

\SetInternalRegister\hbadness{8000} 

\newcommand\doingARLO[2][]{%
  \ifx\mmref\undefined #1\else #2\fi
}

\begin{document}

\title 
      []
      {Results on CP Violation from Belle}


\author{T.E. Browder}{
  address={Department of Physics, University of Hawaii, Honolulu, Hawaii},
  email={teb@phys.hawaii.edu},
  thanks={}
}


\copyrightyear  {2001}

\begin{abstract}
I describe the recent measurement
of the CP violating parameter $\sin 2 \phi_1 =(0.99\pm 0.14\pm 0.05)$
 from the Belle experiment at KEK. 
\end{abstract}

\date{\today}

\maketitle

\section{Introduction}

In 1973, Kobayashi and Maskawa (KM) first proposed a model
where $CP$ violation is
incorporated as an irreducible complex phase in the
weak-interaction quark mixing matrix~\cite{KM}.
The idea was remarkable and daring because
it required the existence of six quarks at a time 
when only the $u$, $d$ and $s$ quarks were known to exist,  
The subsequent
discoveries of the $c$, $b$ and $t$ quarks, and the compatibility
of the model with the $CP$ violation observed in the neutral kaon
system led to the 
incorporation of the KM mechanism into the Standard Model,
even though it had not been conclusively tested experimentally.

In 1981, Sanda, Bigi and Carter pointed out that 
a consequence of the KM model was that large $CP$ violating
asymmetries could occur
in certain decay modes of the $B$ mesons.\cite{carter} 
These asymmetries may occur when a neutral $B$ meson
decays to a $CP$ eigenstate. In this case the amplitude 
for the direct decay interferes with that for the
process where the $B$ meson first mixes to a $\overline{B}$
meson that then decays to the $CP$ eigenstate.

The unitarity of the CKM matrix implies that the existence of three
measurable phases. In the ``Nihongo'' convention, these are denoted
\begin{equation}
\phi_1\equiv arg
\left( \begin{array}{c}
-\frac{V_{cd}V^*_{cb}}{V_{td}V^*_{tb}}
\end{array} \right) 
,~~ \phi_2\equiv arg
\left( \begin{array}{c}
-\frac{V_{ud}V^*_{ub}}{V_{td}V^*_{tb}}
\end{array} \right)
,~~ \phi_3\equiv arg
\left( \begin{array}{c}
-\frac{V_{cd}V^*_{cb}}{V_{ud}V^*_{ub}}
\end{array} \right).
\end{equation}
while at SLAC these angles are usually 
referred to as $\beta, \alpha$ and $\gamma$, respectively.

A non-zero value of $\phi_{CP}$ results in 
the time dependent asymmetry 
\begin{equation}
 A_f =\frac{R(B^0\to f_{CP} )-R(\bar{B}^0\to f_{CP})}
{R(B^0\to f_{CP} )+R(\bar{B}^0\to f_{CP} )} 
=\xi_f \sin 2\phi_{CP} \cdot \sin (\Delta m \cdot (t_2 \pm t_1) ), 
\end{equation}
where $\xi_f$ is the CP eigenvalue ($\pm 1$),
$\Delta m$ denotes the mass difference between the two $B^0$ mass
eigenstates and $t_1$ and $t_2$ are the proper time
for the tagged-$B$ and $CP$ eigenstate decays, respectively. 
The $+$ sign corresponds to the case where the $B^0$ and
$\overline{B^0}$ 
are in an even L orbital angular momentum state, the $-$ sign 
obtains for odd L states such as the $\Upsilon(4S)$.
A determination of $A_f$ thus provides
a measurement of $\sin 2\phi_{CP}.$

We note that due to the restrictions of quantum mechanics,
time integrated asymmetries at the $\Upsilon(4S)$ resonance 
(which corresponds to the $-$ sign in the above equation) are
identically zero. 
Therefore, one must make time dependent measurements. Since
the  pairs of B mesons are produced nearly at rest in the usual
arrangement
at threshold, the $\Upsilon(4S)$ center of mass frame
must be boosted. This is accomplished by the use of beams with
asymmetric energies.
For example at KEK-B, $\beta\gamma\sim 0.43$, 
and as a result the typical $B$
meson decay length is dilated from $20\mu$m
to  about $200 \mu$m, which is measurable with double-sided silicon
strip vertex detectors close to the interaction point.

The measurement therefore requires:
\begin{itemize}
\item
a large sample of reconstructed $B\to (c\overline{c})K^0$ eigenstate
decays;
\item
a determination of the flavor of the accompanying $B$ (``tagging'');
\item 
a measurement of $\Delta z$, the vertex separation between the
$CP$ eigenstate and flavor tag decays;  and
\item
a fit to the flavor-tagged vertex distribution to extract $\sin 2\phi_1$.
\end{itemize}

The KEKB high luminosity double storage ring facility
was commissioned with remarkable speed starting in late 1998.
This accelerator facility allows us to satisfy the first
requirement. By the summer of 2001, $29.1$ fb$^{-1}$ was integrated
on the $\Upsilon(4S)$. This data sample was used for the
$\sin 2\phi_1$ measurement and corresponds to 31.3 million
$B\bar{B}$ pairs. KEKB
uses a $\pm 11$ mrad crossing angle to separate the incoming
and outgoing beams and minimize parasitic collisions. So far
no special limitations associated with
 the crossing angle have been observed (e.g. synchrobetatron oscillations). 
KEKB now routinely achieves peak instantaneous 
luminosities above $5\times 10^{33}/cm^2/sec$ with acceptable 
experimental backgrounds and trigger rates in the Belle detector.
The beam currents are still far below the design values 
and therefore there is room for further improvements in luminosity.
Much larger data samples are expected in the near future.

The Belle detector has good lepton identification
and high efficiency for both charged and neutral particles.
It allows the CP eigenstate decays to be efficiently reconstructed.
Belle is a large-solid-angle magnetic
spectrometer that
consists of a three-layer silicon vertex detector (SVD),
a 50-layer central drift chamber (CDC), a mosaic of
aerogel threshold \v{C}erenkov counters (ACC), time-of-flight
scintillation counters (TOF), and an array of CsI(Tl) crystals
(ECL)  located inside 
a superconducting solenoid coil that provides a 1.5~T
magnetic field.  An iron flux-return located outside of
the coil is instrumented to identify
muons and $K_L$'s (KLM).  
The detector is described in detail elsewhere~\cite{Belle_nim}.
Examples of its performance are given in the following section.

The Belle experiment starting physics data taking in 1999.
In the summer of 2001, Belle (together with BABAR)
announced the observation of the
first statistically significant signals for CP violation 
outside of the kaon system. In this report I will describe
some of the details of the measurement.

\section{$\sin 2\phi_1$ measurement}

We reconstruct $B^0$ decays to the following ${CP}$
eigenstates~\cite{CC}:
$J/\psi K_S$, $\psi(2S)K_S$, $\chi_{c1}K_S$, $\eta_c K_S$ for
$\xi_f=-1$  and $J/\psi K_L$ for $\xi_f=+1$.
We also use $B^0\to J/\psi K^{*0}$ decays where
$K^{*0}\to K_S\pi^0$.  
Here the final state is a mixture of even
and odd $CP$, depending on the relative orbital angular momentum of
the $J/\psi$ and $K^{*0}$.
The $CP$ content is determined from a fit to the full angular
distribution of all $J/\psi K^*$ decay modes other 
than $K^{*0}\to K_S\pi^0$.
We find that the final state is primarily $\xi_f=+1$;
the $\xi_f = -1$ fraction is $0.19 \pm 0.04({\rm stat})\pm 
0.04({\rm syst})$~\cite{Itoh}.

$J/\psi$ and  $\psi(2S)$ mesons are reconstructed via their decays to
$\ell^+\ell^-$ $(\ell=\mu,e)$.
The $\psi(2S)$ is also reconstructed via $J/\psi\pi^+\pi^-$,
and the $\chi_{c1}$ via $J/\psi\gamma$.  The
$\eta_c$ is detected in the $K^+K^-\pi^0$ and
$K_S K^-\pi^+$ modes.
For the $J/\psi K_S$ mode, we use $K_S\to \pi^+\pi^-$ and
$\pi^0\pi^0$
decays; for other modes we only use $K_S\to \pi^+\pi^-$.

The $J/\psi$ and $\psi(2S)\to\mu^+\mu^-$ candidates are
reconstructed from
oppositely charged track pairs where at least
one track is positively identified as a muon by the KLM system 
and the other is either positively identified as a muon or has 
an ECL energy deposit consistent with that of a minimum ionizing
particle. For $e^+e^-$ decays, we use oppositely charged 
track pairs  where at least one track is a well identified
electron and the other track satisfies 
minimal $dE/dx$ and $E/p$ requirements.
For dielectrons, we correct for 
final state radiation or bremsstrahlung and recover additional 
$\psi$ candidates by including the four-momentum of every photon 
detected within 0.05 radians of the
original $e^+$ or $e^-$ direction in the $e^+e^-$ invariant mass
calculation.
Candidate $K_S\to \pi^+\pi^-$ decays are oppositely 
charged track pairs that have an invariant mass 
between $482$~and $514~{\rm MeV}/c^2$, which
corresponds to $\pm 3\sigma$ around the $K_S$ mass peak.


\begin{figure}
  \includegraphics[height=.3\textheight]{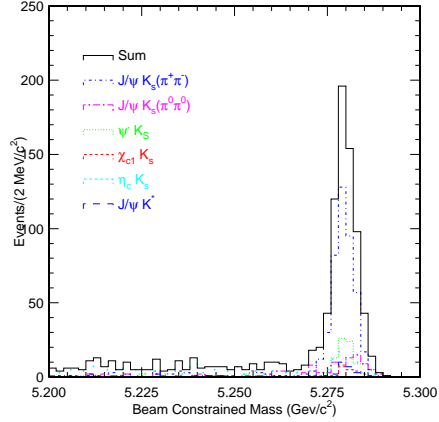}
\caption{The beam-energy constrained mass distribution for
all decay modes combined other than $J/\psi K_L$.
The shaded area is the estimated background.
The signal region is the range $5.27-5.29$ GeV/$\rm{c}^2$.}
\label{fig:belle_mbc}
\end{figure}

To reconstruct $\chi_{c1} K_S$ decays, we select $\chi_{c1}\to  J/\psi\gamma$
decays, rejecting $\gamma$'s that are
consistent with $\pi^0\to \gamma\gamma$
decays, and impose the requirement
$385 < M_{\gamma\ell\ell} - M_{\ell\ell} <430.5~{\rm MeV}/c^2$.
For $\eta_c$ decays, we distinguish kaons from pions
using a combination of CDC dE/dx measurements and information from
the TOF and ACC systems.
Candidate  $\eta_c \rightarrow K^+K^-\pi^0$ $(K_SK^-\pi^+)$ 
decays are selected
with a $K K\pi$ mass requirement that
takes into account the natural width of the $\eta_c$.
For $J/\psi K^{*0}(K_S\pi^0)$ decays, we use $K_S\pi^0$ combinations 
that have an invariant mass 
within 75~MeV/$c^2$ of the nominal $K^*$ mass.
We reduce background from low-momentum $\pi^0$'s by
requiring $\cos\theta_{K^*}<0.8$, where $\theta_{K^*}$ is 
the angle between the $K_S$  momentum vector
and the $K^{*0}$ flight direction calculated  in the $K^{*0}$ rest
frame.

Reconstructed $B$ meson
decays are identified using the beam-constrained 
mass $M_{beam}\equiv\sqrt{E_{beam}^2-p_B^2}$ and
the energy difference $\Delta E\equiv E_B - E_{beam}$,
where $E_{beam}$ is the cms beam energy,
and $p_B$ and $E_B$ are the $B$ candidate 
three-momentum and energy calculated 
in the cms.

Figure~\ref{fig:belle_mbc} shows the combined $M_{\rm bc}$
distribution for
all channels other than $J/\psi K_L$ after
a mode-dependent requirement on $\Delta E$.
The $B$ meson signal region is defined as
$5.270<M_{\rm bc}<5.290~{\rm GeV}/c^2$.
Table~\ref{tab:tally} lists
the numbers of  observed candidates ($N_{\rm ev}$) and
the background ($N_{\rm bkgd}$) determined by
extrapolating the rate from the
 $\Delta E$ {\em vs.} $M_{\rm bc}$ sideband region
into the signal region. About 65\% of the exclusive CP signal
events (44\% of the full CP sample)
are reconstructd in the $B^0\to \psi K_S$, $K_S\to \pi^+\pi^-$ mode.

\begin{table}
\caption{The numbers of  observed
events ($N_{\rm ev}$) and the estimated 
background ($N_{\rm bkgd}$)
in the signal region for each $f_{CP}$ mode.}
\label{tab:tally}
\begin{tabular}{lrr}
Mode & $N_{\rm ev}$ & $N_{\rm bkgd}$\\
\hline
$J/\psi(\ell^+\ell^-) K_S(\pi^+\pi^-)$ & 457 & 11.9\\
$J/\psi(\ell^+\ell^-) K_S(\pi^0\pi^0)$  & 76 & 9.4\\
$\psi(2S)(\ell^+\ell^-)K_S(\pi^+\pi^-)$  & 39 & 1.2\\
$\psi(2S)(J/\psi\pi^+\pi^-)K_S(\pi^+\pi^-)$ & 46 & 2.1\\
$\chi_{c1}(J/\psi\gamma) K_S(\pi^+\pi^-)$ & 24 & 2.4\\
$\eta_c(K^+K^-\pi^0)K_S(\pi^+\pi^-)$ & 23 & 11.3\\
$\eta_c(K_S K^-\pi^+)K_S(\pi^+\pi^-)$ &41 & 13.6\\
$J/\psi K^{*0}(K_S\pi^0)$& 41 & 6.7\\
\hline
Sub-total & 747 & 58.6  \\
\hline
$J/\psi(\ell^+\ell^-) K_L$ & 569 & 223
\end{tabular}
\end{table}

Candidate $B^0\to J/\psi K_L$  decays are selected by requiring
ECL and/or KLM hit patterns that are consistent with the 
presence of a shower induced by a neutral hadron. 
$K_L$ candidates with ECL information only are treated separately
from the $K_L$ candidates with KLM hits.
The centroid of the shower is required to be within a 45$^\circ$ cone
centered on the $K_L$ direction that is inferred from two-body decay
kinematics and the measured four-momentum of the $J/\psi$.
To reduce the background we cut on a likelihood ratio
that depends on the $J/\psi$ cms momentum,
the angle between the $K_L$ and its nearest-neighbor charged track,
the charged track multiplicity of the event, the extent to which the
event is consistent with a $B^+ \to$ $J/\psi K^{*+}(K_L\pi^+)$ hypothesis,
and the polar angle with respect to the $z$ direction
of the reconstructed $B^0$ meson in the cms.
In addition, events that were reconstructed as
$B^0 \to J/\psi K_S$, $J/\psi K^{*0}(K^+\pi^-, K_S \pi^0)$,
$B^+\to$ $J/\psi K^+$, or  $J/\psi K^{*+}(K^+ \pi^0$, $K_S \pi^+)$
decays are removed.  
Finally, $K_L$ clusters with positions that match photons
from reconstructed $\pi^0$'s are also rejected.

\begin{figure}[ht]
  \includegraphics[height=.3\textheight]{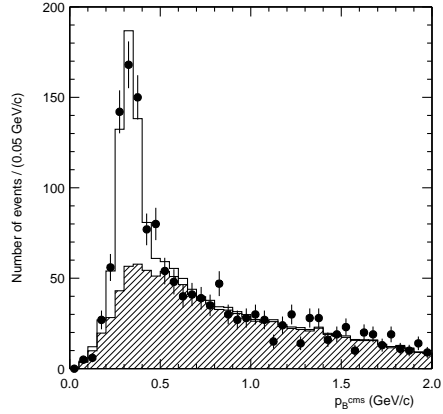}
\caption{The $p_B^{\rm cms}$ distribution for $B^0\to J/\psi K_L$
  candidates
with the results of the fit.
The solid line is the signal plus background;
the shaded area is background only.
The signal region for KLM (ECL-only) clusters
is $0.2\leq p_B^{\rm cms}\leq 0.45 (0.40)~~{\rm GeV}/c$.}
\label{fig:belle_psikl}
\end{figure}

Figure~\ref{fig:belle_psikl} shows the $p_B^{\rm cms}$ distribution,
calculated with the $B^0 \to J/\psi K_L$ two-body decay hypothesis.
The histograms are the results of a fit to the signal
and background distributions.  The shapes are
derived from the Belle GEANT based Monte Carlo (MC) simulation.
However, the normalization and peak position of the signal 
are allowed to vary. There are 397 entries in the
$0.2\leq p_B^{\rm cms}\leq 0.45~~{\rm GeV}/c$
signal region with KLM clusters. 
There are 172 entries in the 
range $0.2\leq p_B^{\rm cms}\leq 0.40~~{\rm
GeV}/c$ with clusters in the ECL only.
The fit finds a total of $346\pm 29$~$J/\psi K_L$ signal events,
and a signal purity of 61\%. Thus, about 33\% of the signal in
the full CP eigenstate sample is reconstructed in the 
$B^0\to \psi K_L$ mode.

To identify
the flavor of the accompanying $B$ meson,
leptons, kaons, $\Lambda$'s, charged slow pions from $D^*\to D^0\pi^+$ decays, 
and energetic pions from two-body $B$ decay (e.g. $\bar{B^0}\to
D^{*+}\pi^+$) are used. A likelihood based method, described
in detail below, is used
to combine information from the different categories and
to take into account their correlations.
The figure of merit for flavor
tagging performance is the effective efficiency, $\epsilon_{eff}$, 
which is $\epsilon(1-2 w^2)$ summed over all tagging categories.
This method gives 
$\epsilon_{eff} = 0.270\pm 0.008^{+0.006}_{-0.009}$.

A Monte Carlo simulation is used to determine a table for a
category-dependent
variable that indicates whether a particle originates from a $B^0$
or $\overline{B}{}^0$.  The values of this variable range from $-1$
for a reliably identified $\overline{B}{}^0$ to $+1$ for a reliably
identified  $B^0$ and  depend on the tagging particle's  charge,
cm momentum,  polar angle, particle-identification probability,
as well as other kinematic and event shape quantities. 
For lepton tags, the missing momentum
and recoil momentum are included in the likelihood determination.
For slow pion tags, the angle between the pion and the thrust axis
of the non-$f_{CP}$ tracks is used. Charged kaon tags
accompanied by $K_S$ mesons,
which have additional strange quark content and a lower
tagging value, are treated as a separate tagging category.

\begin{table}
\caption{The event fractions ($f_l$)
and incorrect flavor assignment probabilities ($w_l$)
for each $r$ interval. The errors include both statistical
and systematic uncertainties.}
\label{tab:tag}
\begin{tabular}{lccc}
$l$&$r$ & $f_l$ & $w_l$\\
\hline
1&$0.000-0.250$ & $0.405$ & $0.465^{+0.010}_{-0.009}$ \\
2&$0.250-0.500$ & $0.149$ & $0.352^{+0.015}_{-0.014}$ \\
3&$0.500-0.625$ & $0.081$ & $0.243^{+0.021}_{-0.030}$\\
4&$0.625-0.750$ & $0.099$ & $0.176^{+0.022}_{-0.017}$\\
5&$0.750-0.875$ & $0.123$ & $0.110^{+0.022}_{-0.014}$\\
6&$0.875-1.000$ & $0.140$ & $0.041^{+0.011}_{-0.010}$
\end{tabular}
\end{table}

The results
from the separate, particle-level categories are then combined
in a second stage that takes correlations for the case of multiple
particle-level tags into account.  This second stage determines
two event-level parameters, $q$ and $r$.  The first, $q$, has the
discrete values $q = +1$  when the tagged $B$ meson is more likely
to be a $B^0$ and $-1$ when it is more likely to be a
$\overline{B}{}^0$.
The parameter $r$ is an event-by-event flavor-tagging dilution
factor
that ranges from $r=0$ for no flavor discrimination to $r=1$ for
unambiguous flavor assignment.  The value of $r$ is used only
to sort data into six intervals of flavor purity;  the wrong-tag
probabilities that are used in the final CP fit are determined from data.

\begin{figure}
  \includegraphics[height=.3\textheight]{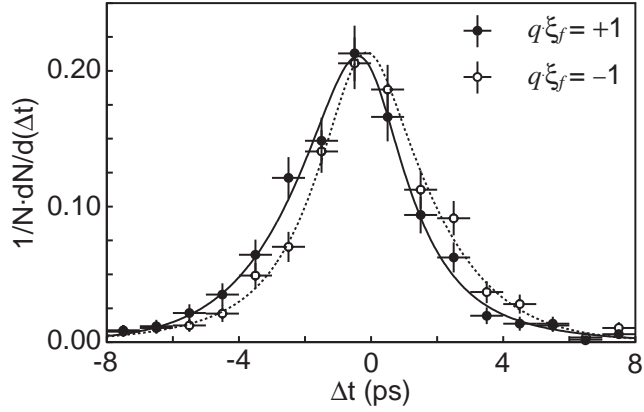}
\caption{$\Delta t$ distributions 
for the events with $q\xi_f = +1$ (solid
points) and $q\xi_f = -1$ (open points). The 
results of the global fit (with  $\sin 2\phi_1 = 0.99$)
are shown as solid and dashed curves, respectively.}
\label{fig:dNdt}
\end{figure}

To avoid dependence on the MC, the probabilities of an
incorrect flavor assignment, $w_l\ (l=1,6)$,
are determined directly from the data for six  $r$ intervals
using exclusively reconstructed, self-tagged
$B^0\to D^{*-}\ell^+\nu$, $D^{(*)-}\pi^+$,  
$D^{*-}\rho^+$  and $J/\psi K^{*0}(K^+\pi^-)$ decays.
The $b$-flavor of the accompanying $B$ meson
is assigned according to the flavor-tagging algorithm described
above. The exclusive decay and tag vertices are reconstructed
using the same vertexing algorithm that is used in
the $CP$ fit. The values of
$w_l$ are obtained from the amplitudes of the
time-dependent $B^0\overline{B}{}^0$ mixing oscillations:
$(N_{\rm OF} - N_{\rm SF})/(N_{\rm OF}+N_{\rm SF})
=(1-2w_l )\cos (\Delta m_d \Delta t)$.
Here $N_{\rm OF}$ and $N_{\rm SF}$ are the numbers of opposite 
flavor and same flavor events.
The value of $\Delta m_d$ is fixed at the world average~\cite{PDG}.
Table~\ref{tab:tag} lists the resulting  $w_l$ values
together with the fraction of the events ($f_l$)
in each $r$ interval.

The $f_{CP}$ vertex is determined using
lepton tracks  from $J/\psi$ or $\psi(2S)$ decays, 
or prompt tracks from $\eta_c$ decays.
The $f_{\rm tag}$ vertex
is determined from well reconstructed tracks not assigned to
$f_{CP}$. Tracks that form a $K_S$ are not used. The tracks
used for vertexing must have at least one three dimensional point with
consistent hits in $r-\phi$ plus at least one additional z hit.
 Each vertex position is required to be consistent with the
interaction point profile smeared in the $r$-$\phi$ plane by the
$B$ meson decay length.  
We use an iterative
procedure:  if the quality of the vertex fit is poor, the track that
is the largest contributor to the $\chi^2$ is removed and the fit is
repeated.
The typical vertex-finding efficiency and
vertex resolution (rms) for $z_{CP}$ ($z_{\rm tag}$) are
$92~(91)\%$ and $75~(140)~\mu{\rm m}$, respectively. 
Note that the resolution on the tag side includes a large 
additional contribution from charm decay.
 Once the tag and CP eigenstate vertices are reconstructed,
the proper time is calculated from $\Delta z/\gamma \beta$.

After vertexing there are  560 events with $q=+1$ flavor tags  and 
577 events with $q=-1$. Figure~\ref{fig:dNdt} shows the  
observed $\Delta t$ distributions
for the $q\xi_f =+1$ 
(solid points) and $q\xi_f = -1$ (open points) event samples. 
In the raw data
there is a clear asymmetry between the two distributions; this 
demonstrates visually  that $CP$ symmetry is violated.

To extract the measured value of the CP violating parameter,
we perform an unbinned maximum likelihood fit to the 
time distributions of the tagged and vertexed events.
The fit takes into account the effects of background, vertex
resolutions and incorrect tagging.

\begin{table}
\caption{The values of
$\sin 2\phi_1$  for various  subsamples
(statistical errors only).}
\label{tab:checks}
\begin{tabular}{ll}
Sample & $\sin 2\phi_1$\\
\hline
$f_{\rm tag}=B^0$ ($q=+1$) &  $0.84\pm 0.21$\\
$f_{\rm tag}=\overline{B}{}^0$ ($q=-1$) & $1.11\pm 0.17$\\
\hline
$J/\psi K_S(\pi^+\pi^-)$ & $0.81\pm 0.20$\\
$(c\bar{c})K_S$ except $J/\psi K_S(\pi^+\pi^-)$ & $1.00 \pm 0.40$\\
$J/\psi K_L$  & $1.31\pm 0.23$\\
$J/\psi K^{*0}(K_S\pi^0)$ & $0.85 \pm 1.45$\\
\hline 
All & $0.99\pm 0.14$
\end{tabular}
\end{table}

For modes other than $J/\psi K^{*0}$
the pdf expected for the signal is
$$
\begin{array}{l}
{\cal P}_{\rm sig}(\Delta t,q,w_l,\xi_f)\\
\quad=\frac{e^{-|\Delta t|/\tau_{B^0}}}{2\tau_{B^0}}
\{1-\xi_f q(1-2w_l)\sin 2\phi_1\sin (\Delta m_d\Delta t )\},
\end{array}
$$
where we fix $\tau_{B^0}$ and $\Delta m_d$ at  their world average
values~\cite{PDG}. For the $B^0\to \psi K^{*0}$ mode 
a more complex fit
to $\Delta t$ and the transversity angle\cite{transversity} is used to take
into account the two CP eigenstates which contribute\cite{Itoh}.

The pdf used for the background distribution is
${\cal P}_{\rm bkg}(\Delta t) = f_\tau e^{-|\Delta t|/\tau_{\rm
bkg}}/2
\tau_{\rm bkg}+(1-f_\tau)\delta(\Delta t),$  where $f_\tau$ is the 
fraction of the background component with an effective lifetime 
$\tau_{\rm bkg}$ and $\delta$ is the Dirac delta function.  For all 
$f_{CP}$ modes other than $J/\psi K_L$, a study using events in the
$\Delta E$ {\em vs.}  $M_{\rm bc}$ sideband regions shows that 
the $f_{\tau}$ component is negligible.   For these modes we use
${\cal P}_{\rm bkg}(\Delta t) = \delta(\Delta t)$.

\begin{figure}
  \includegraphics[height=.3\textheight]{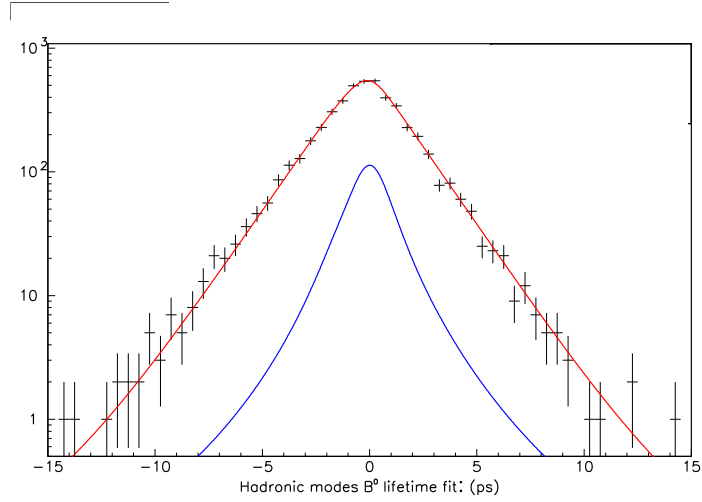}
\caption{The proper decay time distributions for $B^0\to
D^{(*)-}\pi^+$ decays.
The upper curve is the result of the fit for the $B^0$ lifetime; the
lower curve represents the background contribution.}
\label{fig:resol}
\end{figure}

In the case of 
$B^0\to \psi K_L$, the  background is dominated by $B \to \psi X$ decays 
where some final states are $CP$ eigenstates. We estimate the
fractions of the background  components 
with and without a true $K_L$ cluster by 
fitting  the $p_B^{\rm cm}$ distribution to the expected shapes
determined from MC.   
We also use the MC to determine the fraction of events 
with definite $CP$ content within each component.  The result is a 
background that is 71\% from non-$CP$ modes with $\tau_{\rm
bkg}=\tau_B.$  
For the $CP$-mode backgrounds, we use the signal pdf given above
with  the appropriate $\xi_f$ values.   For $\psi K^*(K_L\pi^0)$, which 
is $13\%$ of the background,  we use the $\xi_f=-1$ content
determined from the full $\psi K^*$ sample. The remaining backgrounds are
$\xi_f=-1$ states ($10\%$) including  $\psi K_S$,  and  $\xi_f=+1$ 
states  ($5\%$) including $\psi(2S)K_L$,  $\chi_{c1}K_L$ and 
$\psi \pi^0$.

\begin{figure}
  \includegraphics[height=.6\textheight]{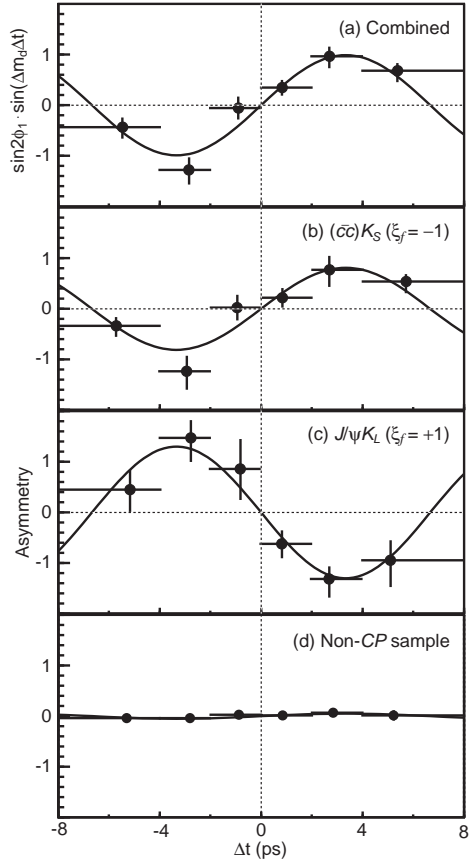}
\caption{(a) The asymmetry obtained
from separate fits to each $\Delta t$ bin for 
the full data sample; the curve is the result of 
the global fit. The
corresponding plots for the (b) $(c\bar{c})K_S$ ($\xi_f=-1$), (c) 
$J/\psi K_L$ ($\xi_f = +1$), and (d) $B^0$ control samples
are also shown.  The curves
are the results of the fit applied separately to the
individual data samples.}
\label{fig:asym}
\end{figure}

The last ingredient needed for the CP fit is the
proper-time interval resolution, $R(\Delta t)$. 
This resolution function is parameterized by 
convolving a sum of two  Gaussians (a {\it main} component due to 
the SVD vertex resolution and charmed meson lifetimes, plus a
{\it tail} component caused by poorly reconstructed tracks) with a 
function that takes into account the cm motion of the $B$ mesons.
The relative fraction of the main Gaussian is determined to be 
$0.97\pm 0.02$ from a study of $B^0\to D^{*-}\pi^+$, $D^{*-}\rho^+$,  
$D^-\pi^+$, $\psi K^{*0}$, $\psi K_S$ and 
$B^+\to \overline{D}{}^0\pi^+$,  $\psi K^+$ events. The means 
($\mu_{\rm main}$, $\mu_{\rm tail}$) and widths ($\sigma_{\rm
main}$, 
$\sigma_{\rm tail}$) of the Gaussians are calculated event-by-event 
from the $f_{CP}$ and $f_{\rm tag}$ vertex-fit error matrices
and the $\chi^2$ values of the fit; typical values are 
$\mu_{\rm main}=-0.24~{\rm ps}$, $\mu_{\rm tail}=0.18~{\rm ps}$
and $\sigma_{\rm main}=1.49~{\rm ps}$, $\sigma_{\rm tail}=3.85~{\rm
ps}$. An example of a fit for hadronic non-CP eigenstate
using the resolution function 
modes is shown in Fig.~\ref{fig:resol}. The fit
agrees well with data out to 10 lifetimes on the logarithmic scale
of the figure. 
As a consistency check, we obtain lifetimes for the neutral and 
charged $B$ mesons using the same vertexing procedure that is used
for the CP fit.
The results of the fit agree well with the world average $B^0$
lifetime value.\cite{Deltam}

The pdfs for signal and background
are convolved with $R(\Delta t)$ to determine the likelihood 
value for each event as a function of $\sin 2\phi_1$:
$$
\begin{array}{lcl}
{\cal L}_i&=&{\displaystyle
\int}\{f_{\rm sig}{\cal P}_{\rm sig}(\Delta t^\prime,
q,w_l,\xi_f) R_{sig}(\Delta t - \Delta t^\prime)\\
&&+(1-f_{\rm sig}){\cal P}_{\rm bkg}(\Delta t^\prime) R_{bkg}(\Delta
t-\Delta
t^\prime)\}d\Delta t^\prime,
\end{array}
$$
where $f_{\rm sig}$ is the probability that the event is signal. The 
most probable $\sin 2\phi_1$ is the value that maximizes the
likelihood 
function $L=\prod_i {\cal L}_i$, where the product is over all
events. (Note that the signal and background resolution functions
are different.)

The result of the fit is 
$$  \sin 2\phi_1 = 
0.99 \pm 0.14 ({\rm stat}) \pm 0.06({\rm syst}).$$
In Fig.~\ref{fig:asym}(a) we show the 
asymmetries for the combined data sample 
that are obtained by applying the fit to the events in each
$\Delta t$ bin separately.  The smooth curve is the result
of the global unbinned fit.
Figures~\ref{fig:asym}(b) and (c) show the 
corresponding asymmetry displays
for the $ (c\bar{c})K_S$  ($\xi_f=-1$)  
and the $J/\psi K_L$ ($\xi_f=+1$) modes separately. 
The observed
asymmetries for the different $CP$ states are indeed opposite,
as expected. 
The curves are the results of unbinned fits
applied separately to the two samples; the resultant
$\sin 2\phi_1$ values are
$0.84 \pm 0.17$(stat) and $1.31 \pm 0.23 $(stat), respectively.

The systematic error is dominated by
uncertainties in the tails of the vertex 
distributions, which contribute $0.04$.
Other significant contributions come from uncertainties (a) in
$w_l~(0.03)$;
(b) in the parameters of the resolution function ($0.02$); and
(c) in the $J/\psi K_L$ background fraction ($0.02$).
The errors introduced by uncertainties in 
$\Delta m_d$ and $\tau_{B^0}$ are negligible.

A number of checks on the measurement were performed.
Table~\ref{tab:checks} lists the results obtained by applying the
same analysis to various subsamples.
All values are statistically consistent with each other.
The result is unchanged if we use the $w_l$'s determined
separately for $f_{\rm tag}=B^0$ and $\overline{B}{}^0$.
Fitting
to the non-$CP$ eigenstate self-tagged modes 
$B^0\to D^{(*)-}\pi^+$, $D^{*-}\rho^+$, $J/\psi 
K^{*0}(K^+\pi^-)$ and $D^{*-}\ell^+\nu$,
where no asymmetry is expected,
yields $0.05 \pm 0.04$.  The asymmetry distribution for
this control sample is shown in Fig.~~\ref{fig:asym}(d).

A single consistent reconstruction and analysis procedure
is used for all data samples. We verify that, 
as expected, the $\sin2\phi_1$ values for
the different run periods are consistent.
With toy Monte Carlo studies we also verify that the 
analysis procedure and the errors 
in the analysis are well behaved when $\sin 2 \phi_1$ is
close to the physical boundary.
As a further check, we used three independent $CP$ fitting programs
and two different algorithms for the $f_{\rm tag}$ vertexing and found
no discrepancy.

Finally, we comment on the possibility of direct $CP$ violation.
The signal pdf for a neutral $B$ meson decaying into a $CP$
eigenstate can be expressed 
in the more general form
$${\cal P}_{sig} (\Delta t) =
 \frac{ e^{-|\Delta t|/\tau_{B^0}} }{2\tau_{B^0}(1+|\lambda|^2)}
\{
 {1+|\lambda|^2 \over 2} + q(1-2w_l) [\xi_f Im\lambda \sin(\Delta
 m_d\Delta t)     - {1-|\lambda|^2 \over 2} \cos(\Delta m_d\Delta t) ]
\}
$$
where $\lambda$ is a complex parameter that depends on both
$B^0 \overline B^0$ mixing and on the amplitudes for $B^0$ and
$\overline B^0$ decay
to a $CP$ eigenstate. The presence of a cosine term ($|\lambda|\ne 1$)
indicates direct $CP$ violation.
For the primary analysis, we assumed $|\lambda| = 1$ 
which is the SM expectation.
In order to test this assumption,
we also performed a fit using the above expression with 
$Im\lambda$ (= ``$\sin 2\phi_1$'') and $|\lambda|$ as free
parameters.
We obtain  
  $|\lambda| = 1.03 \pm 0.09$ and $Im\lambda = 0.99 \pm 0.14$
for all $CP$ modes combined, where the errors are statistical only.
This result confirms the assumption used in our analysis.

\section{Conclusions}

In the summer of 2001, Belle (along with BABAR) presented its
first significant measurement
of the CP violating parameter $\sin 2 \phi_1$. BELLE found
$$ \sin 2 \phi_1 = 0.99 \pm 0.14 \pm 0.06$$ with a statistical
significance of greater than six standard
deviations\cite{belle_prl}. This can be compared to the BABAR
result of $ \sin 2 \phi_1 = 0.59 \pm 0.14 \pm 0.05$\cite{babar_prl}.
The two results
are based on data samples of comparable size (31 million and 32
million $B \bar{B}$ pairs, respectively). 
The efficiencies and
resolutions of the two experiments are also quite similar.
However, although the weighted average of the two results agrees
well with indirect determinations that assume the Standard Model,
the two measurements themselves are only marginally consistent. 
Larger data samples and additional more precise measurements
will be required to fully reconcile these two results for $\sin 2\phi_1$. 
In parallel, a program for the measurement of the other
angles, $\phi_2$ and $\phi_3$, has started.

\begin{theacknowledgments}
I wish to acknowledge the extraordinary achievement of
my colleagues on Belle and the heroic effort of the KEKB accelerator
team. I would also like to thank the conference organizers at
Caltech for their hospitality and for a well-organized
and interesting meeting held under difficult circumstances.
\end{theacknowledgments}


\doingARLO[\bibliographystyle{aipproc}]
          {\ifthenelse{\equal{\AIPcitestyleselect}{num}}
             {\bibliographystyle{arlonum}}
             {\bibliographystyle{arlobib}}
          }

\begin{thebibliography}{99}
\bibitem{CC}
Throughout this report, whenever a mode is quoted the inclusion 
of the charge conjugate mode is implied.
\bibitem{Deltam} The measured
$B$-lifetimes are: $\tau_{B^0} = 1.547 \pm 0.021$~ps and 
$\tau_{B^+} = 1.641 \pm 0.033$~ps (statistical errors only). 
These results are slightly different from recent Belle
lifetime results due to the use
of an improved resolution function.
\bibitem{transversity}
$\theta_{\rm tr}$ is defined as the angle 
between the $\ell^+$ direction in the $J/\psi$ rest frame and the 
$z$-axis,
where the $x$-axis is defined as the direction of motion of the
$J/\psi$ in the $\Upsilon(4S)$
rest frame. The $x$-$y$ plane is defined by the $K^*$ decay
products  in the $J/\psi$ rest frame.
\end{thebibliography}

\end{document}